\documentclass[aps,prl,preprint,groupedaddress,showpacs,amsmath,amssymb]{revtex4}

\usepackage{graphicx}
\usepackage{dcolumn}
\usepackage{bm}

\begin{document}

\title{Six closely related YbT$_2$Zn$_{20}$ (T = Fe, Co, Ru, Rh, Os, Ir) heavy fermion compounds with large local moment degeneracy.}

\author{M. S. Torikachvili,$^{1,2}$ S. Jia,$^1$ E. D. Mun,$^1$ S. T. Hannahs,$^3$ R. C. Black,$^4$
W. K. Neils,$^4$ Dinesh Martien,$^4$ S. L. Bud'ko,$^1$ P. C. Canfield$^1$}

\affiliation{$^1$Ames Laboratory US DOE and Department of Physics and Astronomy, Iowa State University, Ames, IA
50011\\ $^2$Department of Physics, San Diego State University, San Diego, CA 92182\\ $^3$National High Magnetic
Field Laboratory, 1800 E. Paul Dirac Dr., Tallahassee, FL 32310\\ $^4$Quantum Design Inc., 6325 Lusk Blvd., San
Diego, CA 92121}

\date{\today}

\begin{abstract}

Heavy fermion compounds represent one of the most strongly correlated forms of electronic matter and give rise to
low temperature states that range from small moment ordering to exotic superconductivity, both of which are often
in close proximity to quantum critical points.  These strong electronic correlations are associated with the
transfer of entropy from the local moment degrees of freedom to the conduction electrons, and, as such, are
intimately related to the low temperature degeneracy of the (originally) moment bearing ion.  Here we report the
discovery of six closely related Yb-based heavy fermion compounds, YbT$_2$Zn$_{20}$, that are members of the
larger family of dilute rare earth bearing compounds: RT$_2$Zn$_{20}$ (T = Fe, Co, Ru, Rh, Os, Ir).  This
discovery doubles the total number of Yb-based heavy fermion materials.  Given these compounds' dilute nature,
systematic changes in T only weakly perturb the Yb site and allow for insight into the effects of degeneracy on
the thermodynamic and transport properties of these model correlated electron systems.

\end{abstract}



\maketitle

Heavy fermion compounds have been recognized as one of the premier examples of strongly correlated electron
behavior for several decades.  Ce- and U-based heavy fermion compounds have been well studied, and in recent years
a small number of Yb-based heavy fermions have been identified as well.\cite{1}  Unfortunately, in part due to the
somewhat unpredictable nature of $4f$ ion hybridization with the conduction electrons, it has been difficult to
find closely related (e.g. structurally) heavy fermion compounds, other than of the ThCr$_2$Si$_2$ structure,
especially Yb-based ones, that allow for systematic studies of the Yb ion degeneracy.  Part of this difficulty is
associated with the fact that the $4f$ hybridization depends so strongly on the local environment of the rare
earth ion.

Dilute, rare earth (R) bearing, intermetallic compounds are ordered materials with less than 5 atomic percent rare
earth fully occupying a unique crystallographic site. Such materials offer the possibility of investigating the
interaction between conduction electrons and $4f$ electrons in fully ordered compounds for relatively low
concentrations of rare earths.  For the case of R = Yb or Ce these materials offer the possibility of preserving
low temperature, coherent effects while more closely approximating the single ion Kondo impurity limit.  A very
promising example of such compounds is derived from the the family of RT$_2$Zn$_{20}$ \cite{2} (T = transition
metal) which has recently been shown to allow for the tuning of the non-magnetic R = Y and Lu members to
exceedingly close to the Stoner limit as well as allowing for the study of the effects of such a highly
polarizable background on local moment magnetic ordering for R = Gd. \cite{3}

In this letter we present thermodynamic and transport data on six, new, strongly correlated Yb-based intermetallic
compounds found in the RT$_2$Zn$_{20}$ family for T = Fe, Co, Ru, Rh, Os, and Ir, effectively doubling the number
of known Yb-based heavy fermions (compounds with linear coefficient of specific heat, $\gamma$, greater than 400
mJ/mol K$^2$ \cite{1}). The RT$_2$Zn$_{20}$ compounds crystallize in the cubic CeCr$_2$Al$_{20}$ ($Fd\bar{3}m$
space group) structure. \cite{4,5}  Due to the relatively low concentration of rare earth, as well as transition
metal, in these compounds, the four nearest neighbors as well as the twelve next-nearest neighbors of the rare
earth ion are Zn atoms. The rare earth ion is coordinated by a 16 atom Frank - Kasper polyhedron and has a cubic
point symmetry. This near spherical distribution of neighboring Zn atoms gives rise to the possibility of
relatively low crystal-electric-field (CEF) split levels and also promises a large degree of similarity between
this isostructural group of Yb-based heavy fermions. These compounds, then not only greatly expand the number of
known Yb-based heavy fermions, but, as will be shown below, also provide a route to studying how the degeneracy of
the Yb-ion at Kondo temperature, $T_K$, effects the low temperature correlated state.

\begin{figure}
\begin{center}
\includegraphics[angle=0,width=150mm]{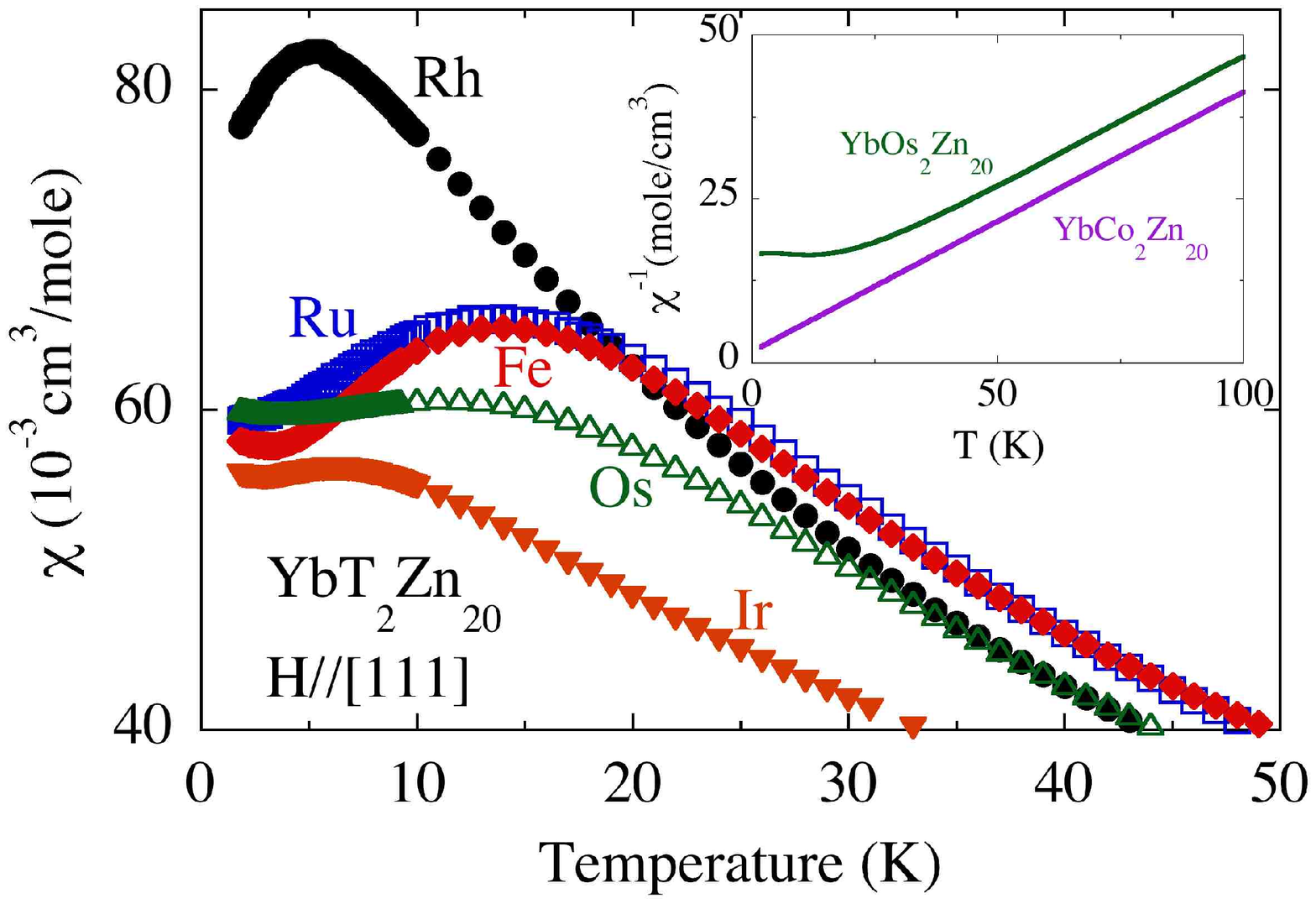}
\includegraphics[angle=0,width=150mm]{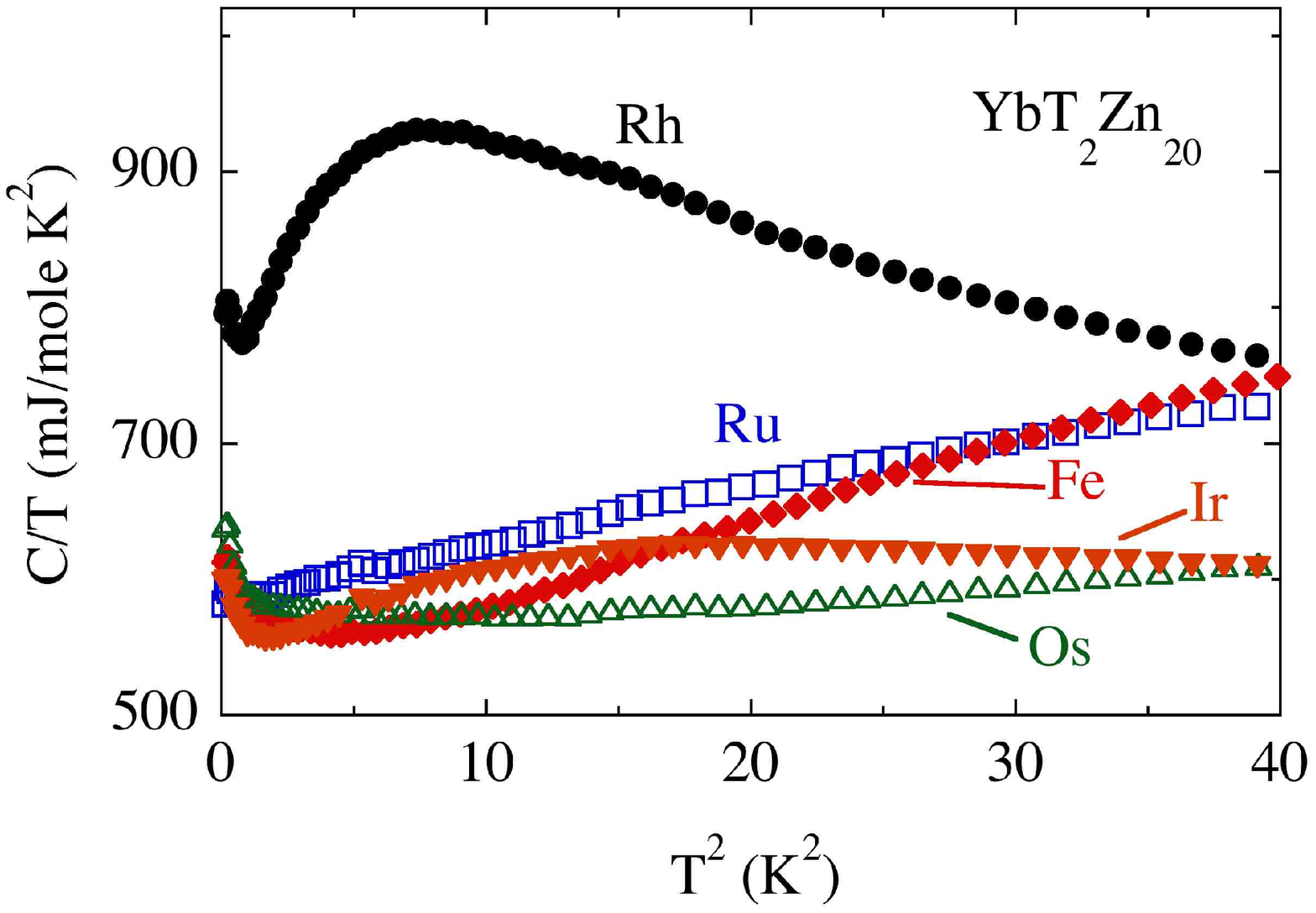}
\end{center}
\caption{{\bf Low temperature thermodynamic properties of YbT$_2$Zn$_{20}$ compounds (T = Fe, Ru, Rh, Os, Ir):}
(a) Magnetic susceptibility ($H = 0.1$ T) . Inset:  Temperature dependent inverse susceptibility for
YbCo$_2$Zn$_{20}$ and YbOs$_2$Zn$_{20}$. (b)Low temperature specific heat, $C$, divided by temperature, as a
function of $T^2$.}\label{F1}
\end{figure}

\begin{figure}
\begin{center}
\includegraphics[angle=0,width=150mm]{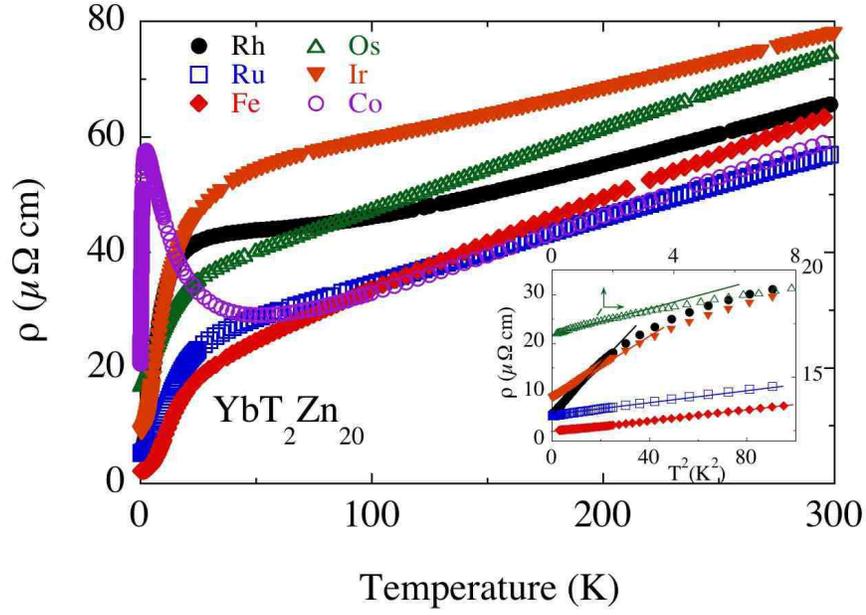}
\end{center}
\caption{{\bf Temperature dependent electrical resistivity of YbT$_2$Zn$_{20}$ compounds (T = Fe, Co, Ru, Rh, Os,
Ir).} Inset: Low temperature electrical resistivity as a function of $T^2$ for T = Fe, Ru, Rh, Os, Ir; note
separate axes for T = Os on top and right.}\label{F3}
\end{figure}

\begin{figure}
\begin{center}
\includegraphics[angle=0,width=150mm]{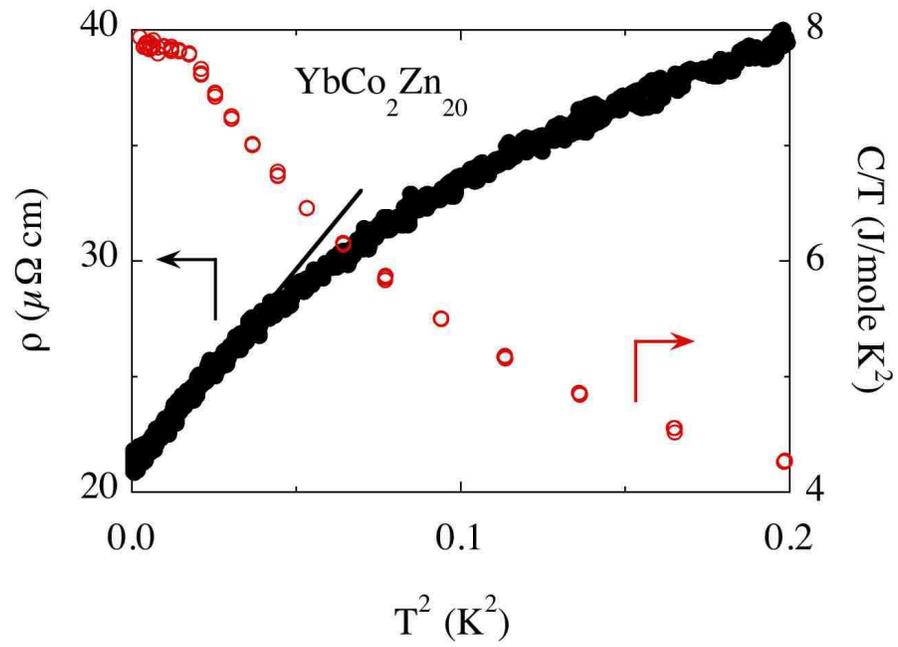}
\end{center}
\caption{{\bf Low temperature electrical resistivity and  {\it C/T} of YbCo$_2$Zn$_{20}$ as a function of {\it
T$^2$}.}}\label{F4}
\end{figure}

Thermodynamic and transport data taken on the six YbT$_2$Zn$_{20}$ compounds are presented in figures
\ref{F1}-\ref{F4} and are summarized in table 1.  Whereas the temperature dependent magnetic susceptibility,
electrical resistivity and specific heat for T = Fe, Ru, Rh, Os, Ir are qualitatively similar, YbCo$_2$Zn$_{20}$
is, at first glance, somewhat different. Most conspicuously, instead of manifesting a clear loss of local moment
behavior at low temperature,\cite{7} the temperature dependent susceptibility continues to be Curie-Weiss-like
down to 2 K (fig. \ref{F1}a, inset).

Focusing initially on the five, apparently similar YbT$_2$Zn$_{20}$ compounds (for T = Fe, Ru, Rh, Os, Ir),
figures \ref{F1}a,b demonstrate that each of these compounds appears to be an excellent example of a Yb-based
heavy fermion with electronic specific heat, $\gamma$, values ranging between 500 and 800 mJ/mole K$^2$. (The
modest rise in the $\Delta C(T)/T$ data below 2 K is most probably associated with a nuclear Schottky anomaly and,
for this work, is simply ignored. This assumption is further supported by the data and analysis presented in Fig.
\ref{ra} below.) The low temperature magnetic susceptibility correlates well with the electronic specific heat
values leading to the Wilson ratio (WR) \cite{1} for these five compounds having values of 1.1 and 1.3 (see table
1). The temperature dependent electrical resistivity data (fig. \ref{F3}) for these five compounds are also
remarkably similar at high temperature and manifest clear $T^2$ temperature dependencies at low temperatures (see
inset). Although resistivity data were taken down to 20 mK, no indications of either magnetic order or
superconductivity were found for any of the YbT$_2$Zn$_{20}$ compounds.

The thermodynamic and transport properties of YbCo$_2$Zn$_{20}$ are somewhat different from the other five
compounds.  YbCo$_2$Zn$_{20}$ does not manifest the clear loss of local moment behavior, above 1.8 K, in the
susceptibility data (see fig. \ref{F1}a, inset) and the electrical resistivity and the specific heat only manifest
Fermi liquid behavior for $T \leq 0.2$ K (fig. \ref{F4}).  Although the higher temperature electrical resistivity
of YbCo$_2$Zn$_{20}$ is similar to the other five YbT$_2$Zn$_{20}$ compounds, it manifests a much clearer example
of a resistance minimum and lower temperature coherence peak.

\begin{figure}
\begin{center}
\includegraphics[angle=0,width=150mm]{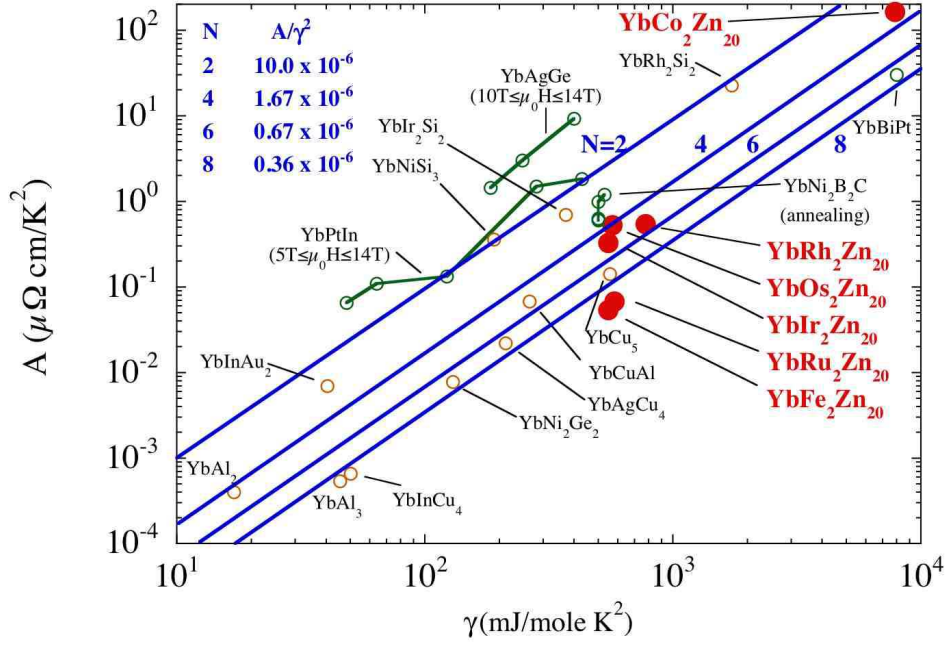}
\end{center}
\caption{{\bf Log-log plot of $A$ versus $\gamma$ (Kadowaki-Woods plot) of six new YbT$_2$Zn$_{20}$ heavy fermion
compounds (T = Fe, Co, Ru, Rh, Os, Ir)} shown with representative data from \cite{10} as well as data for
YbBiPt,\cite{11,12} YbNi$_2$B$_2$C,\cite{14} YbPtIn,\cite{15} YbAgGe,\cite{16} YbNiSi$_3$,\cite{17} and
YbIr$_2$Si$_2$.\cite{Ir} The solid lines for degeneracies N = 2, 4, 6, and 8 are taken from ref.
\cite{10}.}\label{F5}
\end{figure}

Some of the salient parameters extracted from these data are summarized in Table 1 and the coefficient of the
$T^2$ resistivity ($A$) is plotted as a function of the linear coefficient of the specific heat ($\gamma$) in a
Kadowaki-Woods (KW) \cite{kw,mi} type plot (fig. \ref{F5}). Perhaps the most noteworthy point that becomes clear
from this presentation of the data is that, whereas there is relatively little variation in the low temperature
thermodynamic properties, or Wilson ratio, associated with the T = Fe, Ru, Rh, Os, Ir compounds, there is an order
of magnitude variation in the value of the coefficient of the $T^2$ resistivity, $A$.  This gives rise to a
vertical spread of the KW data points.

Recent theoretical work \cite{10,8,9} has generalized the idea of a fixed KW ratio to one that can vary by over an
order of magnitude, depending upon the value of the degeneracy of the Yb ion when it hybridizes.  Figure \ref{F5}
shows, as solid lines, the four degeneracies possible for the Kramer's, Yb$^{+3}$ ion.  The YbT$_2$Zn$_{20}$ data
indicate that for T = Fe, Ru the Yb ion has a significantly larger degeneracy upon entering the Kondo-screened
state than it does for the T = Rh, Os, Ir compounds.  The data point for YbCo$_2$Zn$_{20}$ approaches the far
extreme of the KW plot, being near to the point associated with the exceptionally heavy fermion, YbBiPt.
\cite{11,12}

As mentioned above, the sole Yb site is one of cubic point symmetery and is surrounded only by Zn in a shell of
very high coordination number. Based on these facts it is anticipated that the Yb ion's Hund's rule, ground state
multiplet will split into a quartet and two doublet states with a small total splitting. If indeed the difference
between YbFe$_2$Zn$_{20}$ and YbRu$_2$Zn$_{20}$ on one hand and YbRh$_2$Zn$_{20}$, YbOs$_2$Zn$_{20}$ and
YbIr$_2$Zn$_{20}$ on the other is the degree to which the Hund's rule ground state degeneracy has been lifted by
crystalline electric field splitting before the Kondo screening takes place, then there should be some indication
of this in other data as well.  If, as Tsujii et al. suggest in ref. \cite{10}, the ratio of $T_K$ to $T_{CEF}$ is
of primary concern, then an examination of figure \ref{F1}a in the light of the Coqblin-Schrieffer model
\cite{7,RH}, specifically figure 1 of ref. \cite{7}, indicates that the larger the ratio of the maximum
susceptibility to the low temperature susceptibility, the larger is the degeneracy that remains in the Yb system
at $T_K$. The ratios of the maximum susceptibility to the low temperature susceptibility for T = Fe and Ru are
1.12 and 1.11 respectively whereas the ratios for T = Rh, Os, Ir are 1.06, 1.01 and 1.01, respectively.  These
values are consistent with a difference in degeneracy of at least $\Delta N =2$ (see Fig. \ref{F5}).

This analysis can be made even more thoroughly by performing a fit \cite{7} to the magnetic component of the
specific heat as well as the magnetic susceptibility over a wide temperature range.  This is shown in figure
\ref{ra} for YbFe$_2$Zn$_{20}$, the compound with the largest degeneracy inferred from the KW plot (Fig. \ref{F5})
as well as from the above analysis. Both data sets are best fit (and {\it very} well fit) by the $J = 7/2$ ($N =
8$) curves. The specific heat data is, in particular, most compelling since the height of the anomaly is not an
adjustable parameter once $N$ is chosen. This analysis further confirms the degeneracy inferred from figure
\ref{F5}. The inset to figure \ref{ra}a shows the magnetic entropy as a function of temperature.  By 60 K it rises
past the $J = 5/2$ value. The fact that it does not reach the $R\ln8$ anticipated is most likely due to (i)
difficulties in accurately modelling the non-magnetic contribution with LuFe$_2$Zn$_{20}$ at high temperatures and
(ii) difficulties associated with taking the difference between two large, comparable values, as well as the fact
that by 60 K a recovery of the full $R \ln8$ is not expected (see fit to $J = 7/2$ in main figure).

Given the above analyses, Figure \ref{F5} can be used to infer approximate degeneracies for the Yb ion in these
YbT$_2$Zn$_{20}$ compounds (see Table \ref{T1}). We can then infer a value of $T_K$ by using $T_K = (R \ln
N)/\gamma$ \cite{fto} or by using $T_K = (N-1)\pi^2Rw_N/3N\gamma$ (where $w_N$ is a multiplicative factor which is
a function of $N$ as discussed in \cite{RH}). These expressions produce $T_K$ values that are within 5\% of each
other for $2 \leq N \leq 8$. It should also be noted that the $T_K$ value estimated by this method is close to
that found by fitting the whole $C_p$ and $\chi$ curves (see Fig. \ref{ra} above). As could be anticipated, $T_K$
values for T = Fe and Ru are indeed larger than those found for T = Rh, Os, Ir.

Given that our earlier work on the RT$_2$Zn$_{20}$ families has shown that T = Fe and Ru compounds manifest
anomalously high temperature, local moment ordering due to the fact that the Y and Lu host materials are close to
the Stoner limit,\cite{3} it is noteworthy that for the YbT$_2$Zn$_{20}$ materials it is the T = Fe and Ru
compounds that appear to be significantly different from the T = Rh, Os, and Ir compounds.  Although we currently
do not have enough data to conclude that this Stoner enhancement of the host material (if it even persists in the
Yb based members) is responsible for the higher $T_K/T_{CEF}$ ratio, such an enhancement certainly could be
responsible for increased $T_K$ values. This question is the focus of an ongoing dilution study.

Although at first glance the data for YbCo$_2$Zn$_{20}$ appear to be different from that of the other members of
this family, at low enough temperatures, it too appears to enter into a Fermi liquid ground state and, as shown in
Fig. \ref{F5}, has an intermediate $N$ value, similar to YbOs$_2$Zn$_{20}$. YbCo$_2$Zn$_{20}$ has a substantially
lower $T_K$, and may be closer to a quantum critical point than the other, T = Fe, Ru, Rh, Os, Ir members of the
family.  If YbCo$_2$Zn$_{20}$ is simply closer to a QCP, then, given that the unit cell dimensions for
YbCo$_2$Zn$_{20}$ are the smallest of the family, this would imply that applications of modest pressure to other
members of the YbT$_2$Zn$_{20}$ family may lead to several new Yb-based compounds for the study of quantum
criticality.

\section*{Methods}
Single crystalline samples of YbT$_2$Zn$_{20}$ were grown out of excess Zn using standard solution growth
techniques. \cite{6} Initial ratios of starting elements (Yb:T:Zn) were 2:4:96 (T = Fe, Co), 2:2:96 (T = Ru, Rh),
1:0.5:98.5 (T = Os), and 0.75:1.5:97.75 (T = Ir).  Crystals were grown by slowly cooling the melt between
$1150^\circ$ C and $600^\circ$ C over approximately 100 hours.  In order to reduce the amount of Zn transported to
the top of the growth ampoule, all growths were sealed under approximately 1/3 atmosphere of high purity Ar and
were also slightly elevated from the hearth plate so as to insure that the top of the ampoule was slightly hotter
than the bottom. Residual Zn flux was etched from the surface of the crystals using diluted HCl (0.5 volume
percent, T = Fe, Co) or acetic acid (1 volume percent, T = Ru, Rh, Os, Ir). As  can be seen in Figure \ref{F1},
there is virtually no low-temperature Curie tail observed in any of the T = Fe, Ru, Rh, Os, Ir compounds,
indicating little, or no local-moment-bearing impurities.

Magnetization measurements were performed for $T \geq 1.8$ K in a Quantum Design MPMS unit with the applied
magnetic field along the [111] crystallographic direction.  Specific heat and transport measurements for $T \geq
0.4$ K were performed in a Quantum Design PPMS system.  Specific heat, $C(T)$, data for 50 mK $\leq T \leq 2$ K
were taken on YbCo$_2$Zn$_{20}$ in a dilution refrigerator insert for the Quantum Design PPMS system. Whereas all
RT$_2$Zn$_{20}$ (R = Yb, Lu, Y; T = Fe, Co, Ru, Rh, Os, Ir) had $\Theta_D$ values near 255 K, the linear component
of the $C(T)$ was low (50 mJ/mol K$^2$ or less) \cite{3} for the Lu- and Y- analogues and greatly enhanced for the
Yb-bearing materials. Transport data were taken for $T$ down to 20 mK at the NHMFL, Tallahassee using an Oxford
dilution refrigerator. Powder X-ray diffraction measurements were performed on a Rigaku Miniflex unit. The
YbT$_2$Zn$_{20}$ (T = Fe, Co, Ru, and Rh) compounds had diffraction patterns and lattice parameters that agreed
well with the data for the RT$_2$Zn$_{20}$ series presented in ref. \cite{2}. Although there are no prior reports
on the ROs$_2$Zn$_{20}$ and RIr$_2$Zn$_{20}$ series, the diffraction patterns for YbOs$_2$Zn$_{20}$ and
YbIr$_2$Zn$_{20}$ were easily indexed to the RT$_2$Zn$_{20}$ structure type. Room temperature unit cell parameters
are given in table 1.

\section*{Acknowledgments}

Ames Laboratory is operated for the U.S. Department of Energy by Iowa State University under Contract No.
W-7405-ENG-82.  This work was supported by the Director for Energy Research, Office of Basic Energy Sciences.
P.C.C. and S.L.B. would like to thank L. McArthur for having introduced them to some of the finer points of QD
options. The support from NSF grant No. DMR-0306165 (MST) is gratefully acknowledged. Work at the NHMFL was
performed under the auspices of the NSF, US DOE, and State of Florida.
\\

Correspondence and requests for materials should be addressed to P.C.C.

\section*{Competing interests statement:}
The authors declare that they have no competing financial interests.

\clearpage

\begin{turnpage}
\begin{table}
\caption{{\bf Summary of structural, thermodynamic and transport data on YbT$_2$Zn$_{20}$ compounds (T = Fe, Co,
Ru, Rh, Os, Ir)} showing cubic lattice parameter, $a$; paramagnetic Curie-Weiss temperature, $\Theta$, and
effective moment, $\mu_{eff}$, obtained from high temperature inverse susceptibility fit; low temperature magnetic
susceptibility, $\chi_0$ taken at 1.8 K; magnetic susceptibility at the maximum, $\chi_{max}$ and corresponding
temperature, $T_{\chi_{max}}$; residual resistivity, $\rho_0$, taken at $T \sim 20$ mK; coefficient of the $T^2$
resistivity, $A$(with range of fit given below); residual resistivity ratio, RRR; linear coefficient of the
specific heat, $\gamma$; Wilson ratio, WR; Kadowaki-Woods ratio, KWR; degeneracy, $N$; and estimated Kondo
temperature, $T_K$. \label{T1}}
\begin{ruledtabular}
\begin{tabular}{c|c|c|c|c|c|c|c|c|c|c|c|c|c|c}
T&$a$&$\Theta$&$\mu_{eff}$&$\chi_0$&$\chi_{max}$&$T_{\chi_{max}}$&$\rho_0$&$A$&RRR&$\gamma$&WR&KWR&$N$&$T_K$ \\
&\AA&K&$\mu_B$&$\frac{10^{-3}cm^3}{mole}$&$\frac{10^{-3}cm^3}{mole}$&K&$\mu\Omega~cm$&$\frac{\mu\Omega~cm}{K^2}$&&$\frac{mJ}{mol~K^2}$&&$\frac{\mu\Omega~cm~mole^2~K^2}{mJ^2}$&&K\\
\hline \hline

Fe&14.062&-24.5&4.8&58.0&65.1&14.0&2.1&$5.4 \cdot 10^{-2}$&31.2&520&1.2&$2.0 \cdot 10^{-7}$&8&33\\
&&&&&&&&($T \le 11$ K)&&&&&&\\

Co&14.005&-5.6&4.5&415.1&&&21&165&2.8&7900&&$27 \cdot 10^{-7}$&4&1.5\\
&&&&&&&&($T \le 0.2$ K)&&&&&&\\

Ru&14.193&-18.5&4.6&58.9&65.4&13.5&5.3&$6.8 \cdot 10^{-2}$&10.9&580&1.1&$2.0 \cdot 10^{-7}$&8&30\\
&&&&&&&&($T \le 11$ K)&&&&&&\\

Rh&14.150&-10.0&4.2&77.7&82.4&5.3&5.6&$54 \cdot 10^{-2}$&11.8&740&1.3&$10.1 \cdot 10^{-7}$&6&20\\
&&&&&&&&($T \le 6$ K)&&&&&&\\

Os&14.205&-17.8&4.5&60.0&60.7&11.5&17&$53 \cdot 10^{-2}$&4.4&580&1.1&$15 \cdot 10^{-7}$&4&20\\
&&&&&&&&($T \le 1$ K)&&&&&&\\

Ir&14.165&-24.2&4.4&55.9&56.3&6.5&8.8&$33 \cdot 10^{-2}$&8.9&540&1.2&$11 \cdot 10^{-7}$&4&21\\
&&&&&&&&($T \le 5$ K)&&&&&&\\

\end{tabular}
\end{ruledtabular}
\end{table}
\end{turnpage}

\clearpage

\clearpage

\clearpage

\clearpage

\clearpage

\begin{figure}
\begin{center}
\includegraphics[angle=0,width=100mm]{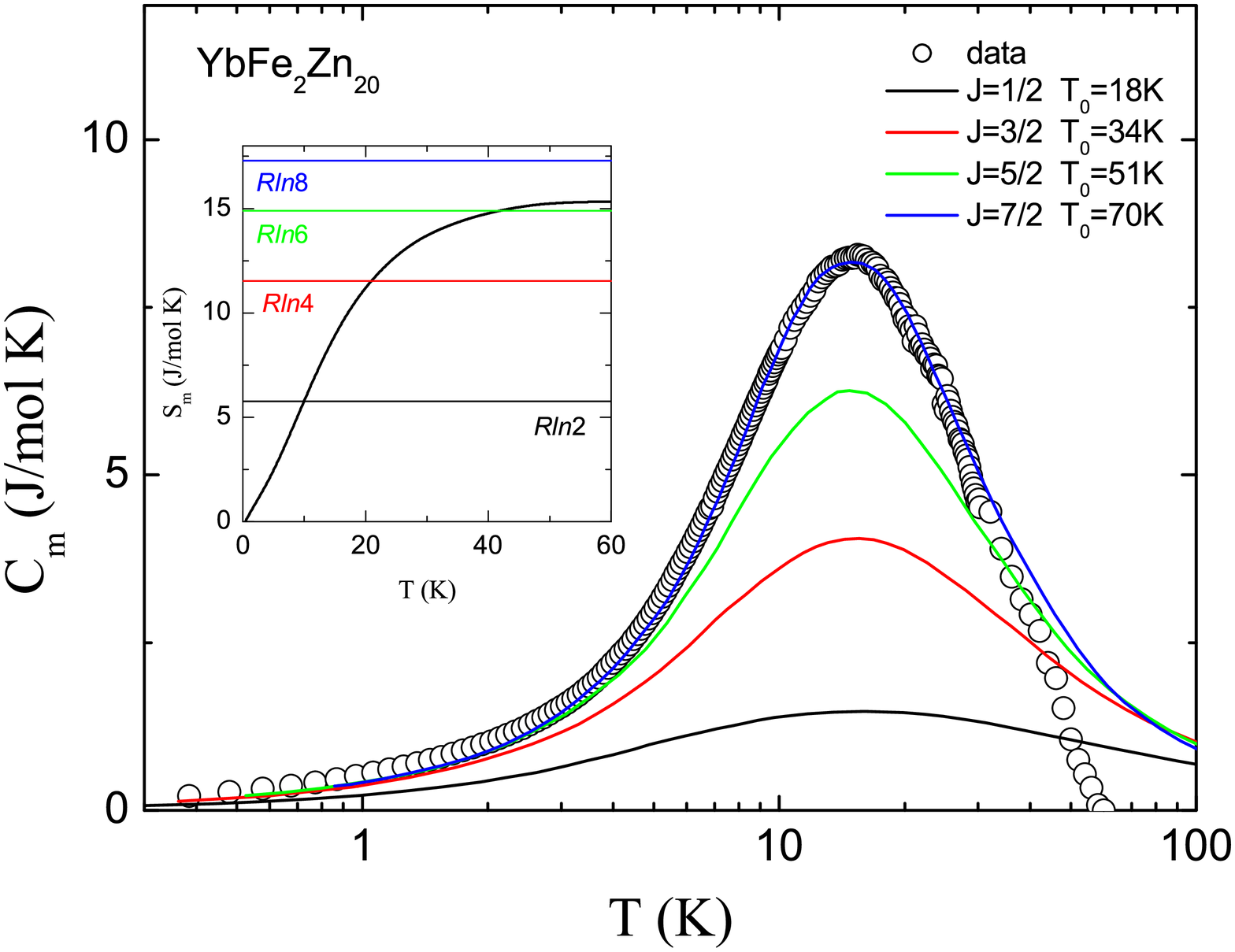}
\includegraphics[angle=0,width=100mm]{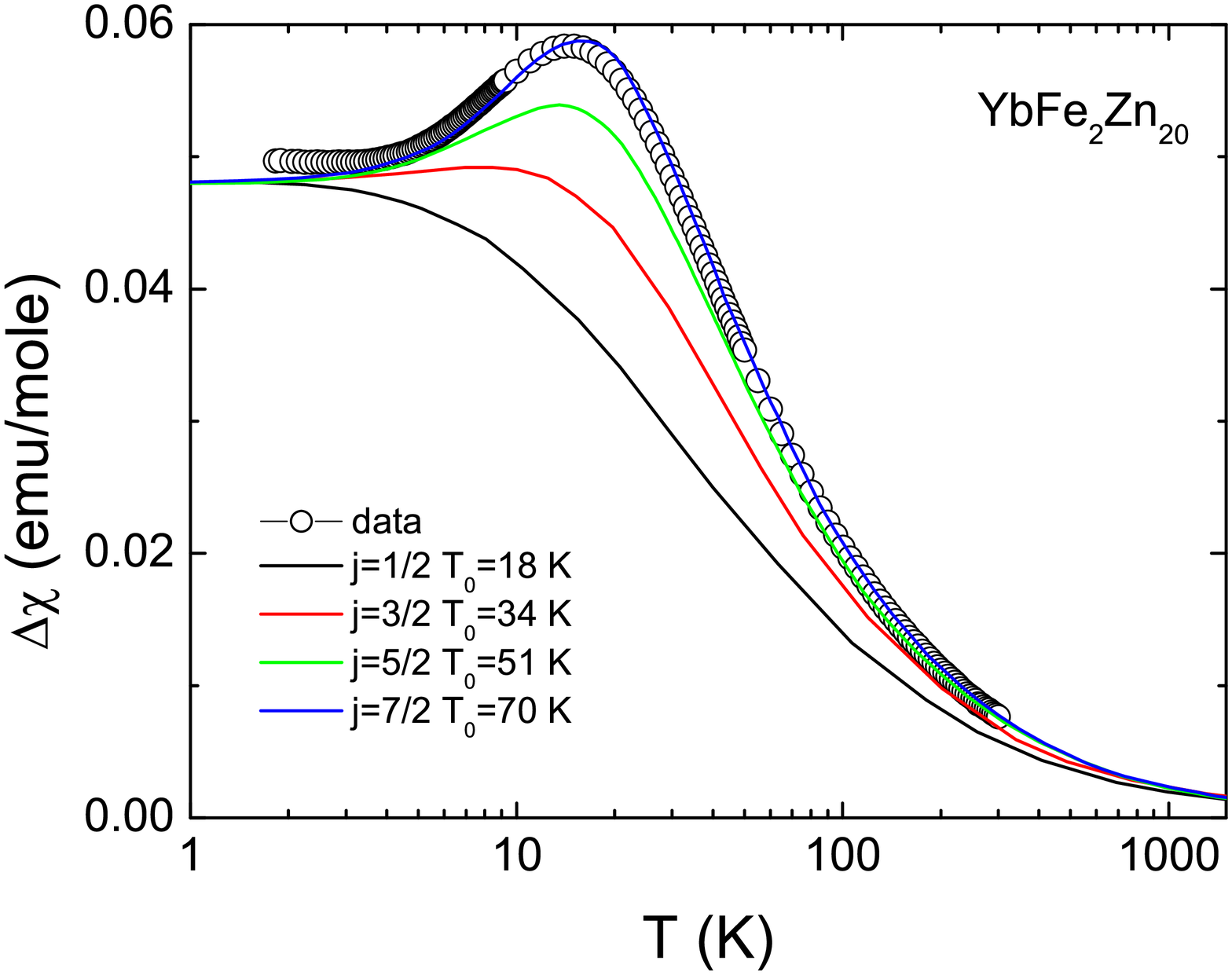}
\end{center}
\caption{{\bf Coqblin-Schrieffer analysis of thermodynamic data from YbFe$_2$Zn$_{20}$ after subtraction data from
the non-magnetic analogue, LuFe$_2$Zn$_{20}$:}  (a) Magnetic specific heat and (b) Net magnetic susceptibility of
YbFe$_2$Zn$_{20}$ as a function of temperature.  Data are shown as open symbol and best fits to $J =
1/2,~3/2,~5/2$ and $7/2$ using formalism described in \cite{7} shown in black, red, green and blue lines
respectively.  $T_0$ values were determined by forcing $C_p$ curves have maxima at the same location as the $C_p$
data.  The Kondo temperature that can be inferred from $T_0 = 70$ K and $J = 7/2$ is $\approx 37$ K. }\label{ra}
\end{figure}


\begin{references}

\bibitem{1}   Stewart, G. R.  Heavy-fermion systems. {\it Rev. Mod. Phys.} {\bf 56}, 755-787 (1984);
Stewart, G. R.  Non-fermi-liquid behavior in d- and f-electron metals. {\it Rev. Mod. Phys.} {\bf 73}, 797-855
(2001) ; Stewart, G. R. Addendum: Non-Fermi-liquid behavior in d- and f-electron metals. {\bf 78,} 743-753 (2006).

\bibitem{2} Nasch,T.,  Jeitschko, W.,  and Rodewald, U. C.  Ternary rare earth transition metal zinc
compounds RT$_2$Zn$_{20}$ with T = Fe, Ru, Co, Rh, and Ni. {\it Zeitschrift f$\ddot{u}$r Naturforschung, B: Chem.
Sciences} {\bf 52}, 1023-1030 (1997).

\bibitem{3} Jia, S.,  Bud'ko, S.L., Samolyuk, G.D., Canfield, P.C. High temperature ferromagnetism in GdFe$_2$Zn$_{20}$:
large, local moments embedded in the nearly ferromagnetic Fermi liquid compound YFe$_2$Zn$_{20}$.
cond-mat/0606615, 1-14 (2006).

\bibitem{4} Kripyakevich, P.I., and Zarechnyuk, O.S. RCr$_2$Al$_{20}$ compounds in systems of rare earth metals and calcium,
and their crystal structures. {\it Dopov. Akad. Nauk Ukr. RSR, Ser. A} {\bf 30}, 364-367 (1968).

\bibitem{5}  Thiede, V.M.T., Jeitschko, W., Niemann, S., and Ebel, T. EuTa$_2$Al$_{20}$, Ca$_6$W$_4$Al$_{43}$ and other
compounds with CeCr$_2$Al$_{20}$ and Ho$_6$Mo$_4$Al$_{43}$ type structures and some magnetic properties of these
compounds. {\it J. Alloys Compd.} {\bf 267}, 23-31 (1998).

\bibitem{7}  Rajan, V.T. Magnetic susceptibility and specific heat of the Coqblin-Schrieffer model. {\it Phys. Rev. Lett.}
{\bf 51}, 308-311 (1983).

\bibitem{kw}  Kadowaki, K., Woods, S.B. Universal relationship of the resistivity and specific heat in heavy-fermion compounds. {\it Sold State Comm.} {\bf 58}, 507-509 (1986).

\bibitem{mi}  Miyake, K., Matsuura, T. Varma, C.M. Relation between resistivity and effective mass in heavy-fermion and A15 compounds. {\it Sold State Comm.}
{\bf 71}, 1149-1153 (1989).

\bibitem{10}  Tsujii, N., Kontani, H., Yoshimura, K. Universality in heavy fermion systems with general degeneracy.
{\it Phys. Rev. Lett.} {\bf 94}, 057201 (2005).

\bibitem{11} Fisk, Z., Canfield, P.C.,  Beyermann, W.P., Thompson, J.D., Hundley, M.F., Ott, H.R., Felder, E.,
Maple, M.B., Lopez de la Torre, M.A. Massive electron state in YbBiPt. {\it Phys. Rev. Lett.} {\bf 67}, 3310-3313
(1991).

\bibitem{12} Movshovich, R., Lacerda, A., Canfield, P.C.,  Thompson, J.D., and Fisk, Z. Low-temperature phase diagram
of YbBiPt. {\it J. Appl. Phys.}, {\bf 76}, 6121-6123 (1994).

\bibitem{14} Avila, M.A., Wu, Y.Q.,  Condron, C.L., Bud'ko, S.L., Kramer, M., Miller, G.J.,  and Canfield, P.C.
Anomalous temperature-dependent transport in YbNi$_2$B$_2$C and its correlation to microstructural features.  {\it
Phys. Rev. B} {\bf 69}, 205107 (2004).

\bibitem{15} Morosan, E., Bud'ko, S.L., Mozharivskyj, Y.A., and Canfield, P.C. Magnetic-field-induced quantum critical
point in YbPtIn and YbPt$_{0.98}$In single crystals. {\it Phys. Rev. B} {\bf 73}, 174432 (2006).

\bibitem{16}  Bud'ko, S.L., Morosan, E., and Canfield, P.C. Magnetic field induced non-Fermi-liquid behavior in
YbAgGe single crystals. {\it Phys. Rev. B} {\bf 69},  014415 (2004).

\bibitem{17} Avila, M.A.,  Sera, M., and Takabatake, T. YbNiSi$_3$: An antiferromagnetic Kondo lattice with strong
exchange interaction. {\it Phys. Rev. B} {\bf 70}, 100409 (2004).

\bibitem{Ir} Hossain, Z.,  Geibel, C., Weickert, F., Radu, T., Tokiwa, Y., Jeevan, H., Gegenwart, P., and
Steglich, F. Yb-based heavy-fermion metal situated close to a quantum critical point. {\it Phys. Rev. B} {\bf 72},
094411 (2005).

\bibitem{8}  Tsujii, N., Yoshimura, K.,  Kosuge, K. Deviation from the Kadowaki-Woods relation in Yb-based
intermediate-valence systems. {\it J. Phys.: Cond. Mat.} {\bf 15}, 1993-2003 (2003).

\bibitem{9}  Kontani, H. Generalized Kadowaki-Woods relation in heavy fermion systems with orbital degeneracy.
{\it J. Phys. Soc. Jpn.} {\bf 73}, 515-518 (2004).

\bibitem{RH} Rasul, J.W., Hewson, A.C. Bethe ansatz and $1/N$ expansion results for $N$-fold degenerate magnetic impurity models. {\it J. Phys. C: Solid State Phys.} {\bf 17},
2555-2573 (1984).

\bibitem{fto}  Fisk, Z., Thompson, J.D., Ott, H.R. Heavy fermions: new materials.
{\it J. Magn. Magn. Mat.} {\bf 76 \& 77}, 637-641 (1988).

\bibitem{6}  Canfield, P.C. and Fisk, Z. Growth of single crystals from metallic fluxes. {\it Phil. Mag. B} {\bf 65},
1117-1123 (1992).

\end{references}
\end{document}